\documentclass[preprint,3p,twocolumn]{elsarticle}
\usepackage[colorlinks]{hyperref}
\usepackage{tabularx}
\usepackage{graphicx}
\usepackage{amssymb,amsmath,bm,tensor,braket}
\usepackage{xcolor}
\usepackage[varg]{txfonts}
\usepackage{enumerate}
\usepackage{mathtools}
\usepackage[utf8]{inputenc}
\usepackage[normalem]{ulem}
\usepackage{lineno,hyperref}
\usepackage{booktabs}
\usepackage[normalem]{ulem}
\modulolinenumbers[5]
\usepackage{tipa}

\journal{Physics Letters B}

\newcommand{\HW}[2]{#2}     

\begin{document}

\begin{frontmatter}

\title{Bounding the photon mass with cosmological propagation 
\\ of fast radio bursts}

\author[1,2]{Huimei Wang}
\author[3,4]{Xueli Miao}
\author[2,4]{Lijing Shao\corref{cor1}}\ead{lshao@pku.edu.cn}
\cortext[cor1]{Corresponding author}

\address[1]{Department of Astronomy, School of Physics, Peking University,
Beijing 100871, China}
\address[2]{Kavli Institute for Astronomy and Astrophysics, Peking
University, Beijing 100871, China}
\address[3]{School of Physics and State Key Laboratory of Nuclear Physics
and Technology, Peking University, Beijing 100871, China}
\address[4]{National Astronomical Observatories, Chinese Academy of
Sciences, Beijing 100012, China}

\begin{abstract}
Photon is the fundamental quantum of electromagnetic fields, whose mass,
$m_{\gamma}$, should be strictly zero in Maxwell's theory. But not all
theories adopt this hypothesis. If the rest mass of the photon is not zero,
there will be an additional time delay between photons of different
frequencies after they travel through a fixed distance. By analyzing the
time delay, we can measure or constrain the photon mass. Fast radio bursts
(FRBs)---transient radio bursts characterized by millisecond duration and
cosmological propagation---are excellent astrophysical laboratories to
constrain $m_{\gamma}$. In this work we use a catalog of 129 FRBs in a
Bayesian framework to constrain $m_{\gamma}$. As a result, we obtain a new
bound on the photon mass, $m_{\gamma} \leq 3.1\times 10^{-51}\rm\,kg\simeq
1.7 \times 10^{-15}\,eV/c^2$ ($m_{\gamma} \leq 3.9\times 10^{-51}\rm\,kg
\simeq 2.2 \times 10^{-15}\,eV/c^2$) at the $68\%$ $(95\%$) confidence
level. The result represents the best limit purely from kinematic analysis
of light propagation. The bound on the photon mass will be tighter in the
near future with increment in the number of FRBs, more accurate measurement
of the redshift for FRBs, and refinement in the knowledge about the origin
of dispersion measures (DMs).
\end{abstract}


\end{frontmatter}


\section{Introduction}

``Constancy of light speed'' is an important postulate of Einstein's theory of
special relativity~\cite{Einstein:1905ve}, and it is adopted in the general
relativity and quantum field theories. In quantum mechanics, the particle-wave
duality translates the constancy of light speed into the ``masslessness of
photons'', indicating that the rest mass of the photon should be strictly zero. 
Nevertheless, there are still a number of theories which challenge this
postulate, including the famous Maxwell-de Broglie-Proca theory where photons
are massive~\cite{Proca:1900nv, deBroglie:1922zz}. \HW{}{In the Standard Model
of particle physics, the gauge symmetry of quantum electrodynamics prevents
photons from acquiring a non-zero mass. Therefore, confirmation of the
masslessness of photons can also be viewed as a need for a gauge theory of
quantum electrodynamics.}

\HW{}{In the viewpoint of testing fundamental principles and pushing the
boundary of contemporary physical theories to a larger extent, we need to
tighten the empirical limits on the photon mass from various experiments.} A
non-zero mass of the photon can cause a series of experimental or observational
consequences.  There are already many efforts to constrain the rest mass of the
photon~\cite{Lowenthal:1973ka, Tu:2005ge,Okun:2006pn,
Goldhaber:2008xy,Spavieri:2011zz}, such as early experiment in testing the
Coulomb law~\cite{Williams:1971ms}, toroid Cavendish balance~\cite{Lakes:1998mi,
Luo:2003rz}, gravitational deflection of electromagnetic
waves~\cite{Lowenthal:1973ka}, magneto-hydro-dynamic phenomena of the Solar
wind~\cite{Retino:2013gga, Ryutov:1997zz, Ryutov:2007zz}, spindown of
pulsar~\cite{Yang:2017ece}, Jupiter's magnetic field~\cite{Davis:1975mn}, the
mechanical stability of magnetized gas in galaxies~\cite{Chibisov:1976mm},
``black hole bombs''~\cite{Pani:2012vp, Press:1972zz} and so on. These are based
on the {\it dynamics} to test the photon mass, and they depend on specific
massive photon theories. Compared with the {\it dynamic} tests, {\it kinematic}
tests~\cite{RN26, Wu:2016brq, Bonetti:2016cpo, Bonetti:2017pym, Zhang:2016ruz,
Shao:2017tuu, Wei:2016jgc, Wei:2018pyh, Xing:2019geq, Wei:2020wtf} are cleaner
since they do not involve any dynamic hypothesis thus are largely independent of
the underlying massive photon theory. In this paper, we constrain the photon
mass from the propagation of electromagnetic waves of fast radio bursts
(FRBs)~\cite{Lorimer:2007qn}, which is a kinematic method. Due to the
uncertainty principle of quantum mechanics and the finite age of our Universe
($\sim 10^{10}$ years), there is an ultimate lower limit on the photon rest mass
that we in principle could probe. It equals to $m_{\gamma} \approx \hbar /
\Delta t_0 c_0^{2} \approx 10^{-69}\, \mathrm{kg}$, where $\hbar$ is the reduced
Planck constant, $\Delta t_0$ is the age of the Universe, $c_0$ is the limiting
velocity for photons of infinite frequency~\cite{Tu:2005ge,RN55}.

Assuming that photons have a non-zero rest mass, $m_{\gamma}$, according to
Einstein's special relativity, the dispersion equation can be written as,
\begin{equation}\label{Eq:E2}
E^{2}=p^{2} c_{0}^{2}+m_{\gamma}^{2} c_{0}^{4}\,,
\end{equation}
where $p$ is the momentum. We can get the group velocity for a photon with
energy $E=h\nu$ as,
\begin{equation}\label{Eq:velocity}
v \equiv \frac{\partial E}{\partial p}=\frac{p c_{0}}{E} c_{0} \simeq
c_{0}\bigg[1-\frac{1}{2}\Big(\frac{m_{\gamma}
c_{0}^{2}}{h\nu}\Big)^{2}\bigg]\,,
\end{equation}
where $h$ is the Planck constant and $\nu$ is the photon frequency. The
last term only holds when $m_{\gamma} \ll h \nu / c_0^{2} \simeq 7 \times
10^{-42}\left({\nu}/{\mathrm{GHz}}\right)\, \mathrm{kg}$. It is suggested
by Eq.~(\ref{Eq:velocity}) that when the energy is smaller, the relative
modification is larger. We constrain the photon mass by the accumulative
time delay from propagation of light. Since a shorter emitting time-scale,
a longer propagation distance, and a lower energy of the photons can
provide a larger time delay, FRBs have some features that deserve our
attention in a test of such kind~\cite{Wu:2016brq, Bonetti:2016cpo, Bonetti:2017pym,
Shao:2017tuu, Xing:2019geq, Wei:2020wtf}:
\begin{enumerate}[(i)]
	\item FRBs belong to a kind of instantaneous events with a time scale
	in milliseconds;
	\item FRBs are widely considered to be cosmological objects and their
	redshifts can be significant in general;
	\item FRBs are observed at radio frequency which means a lower energy
	for photons, so they can provide a tight bound on $m_{\gamma}$ via
	Eq.~(\ref{Eq:velocity}).
\end{enumerate}
The above features make FRBs a great celestial laboratory to limit
$m_{\gamma}$. Previous bounds on the photon mass from FRBs are collected in
Table~\ref{Table:Previous:bounds}~\cite{Wu:2016brq, Bonetti:2017pym, 
Shao:2017tuu, Wei:2018pyh,Xing:2019geq}.

\begin{table*}[htb]
    \centering
	\caption{A collection of bounds on the photon mass from
	FRBs.\label{Table:Previous:bounds} }
    \begin{tabular}{rcc}
	    \toprule
	    	& FRBs & Bound on the photon mass (kg)\\
	    \midrule
		    Wu et al. (2016)~\cite{Wu:2016brq} &FRB 150418&$5.2\times
		    10^{-50}$\\
	    	Bonetti et al. (2017)~\cite{Bonetti:2017pym} &FRB
	    	121102&$3.9\times 10^{-50}$\\
		    Shao $\&$ Zhang (2017)~\cite{Shao:2017tuu} & A
		    combination of 21 FRBs & $\quad\quad\,\,\,\,8.7\times 10^{-51}$
		    $(68\%)$ \\
	    	& &$\quad\quad\,\,\,\,1.5\times 10^{-50}$ $(95\%)$\\
	    	Xing et al. (2019)~\cite{Xing:2019geq} & subbursts of FRB
	    	121102 &$5.1\times 10^{-51}$\\
			Wei $\&$ Wu (2020)~\cite{Wei:2020wtf} & A combination of
			9 FRBs &$\quad\quad\,\,\,\,7.1\times 10^{-51}$ $(68\%)$\\
	    	& &$\quad\quad\,\,\,\,1.3\times 10^{-50}$ $(95\%)$\\
			This work & A combination of 129 FRBs &
			$\quad\quad\,\,\,\,3.1\times 10^{-51}$ $(68\%)$ \\
			&& $\quad\quad\,\,\,\,3.9\times 10^{-51}$ $(95\%)$\\
    	\bottomrule
    \end{tabular}
\end{table*}

With considerably increased radio data from telescopes such as the Canadian
Hydrogen Intensity Mapping Experiment (CHIME), the Australian Square
Kilometre Array Pathfinder (ASKAP), the Molonglo Observatory Synthesis
Telescope (UTMOST), the Five-hundred-meter Aperture Spherical radio
Telescope (FAST), and the Apertif Radio Transient System installed in the
Westerbork Synthesis Radio Telescope (WSRT), the observation of FRBs has
developed a lot in recent years~\cite{Platts:2018hiy}. Therefore, it is
timely to use the most recent data to put an updated bound on the photon
mass.

\begin{figure*}[h]
	\centering
	\includegraphics[width=11cm]{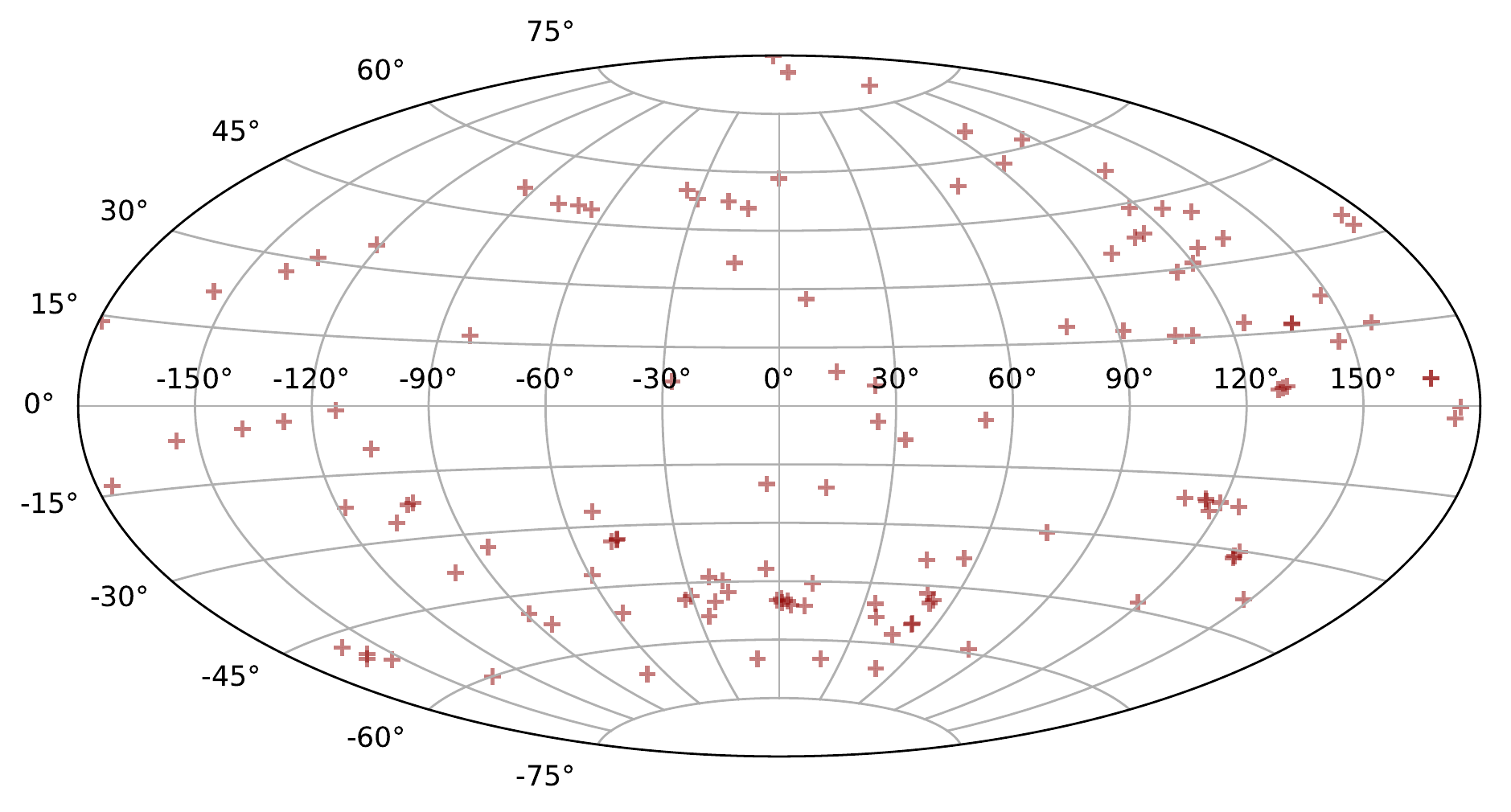}
	\caption{Sky position (in Galactic coordinate) of 129 FRBs which are used in bounding the photon mass.\label{Fig:Distribution}}
\end{figure*}

In this paper, we collect a catalog of FRBs to bound the photon mass.
Figure~\ref{Fig:Distribution} shows the sky distribution of FRBs which are
used in this
work~\cite{Petroff:2016tcr}.\footnote{\url{http://www.frbcat.org} and
\url{https://www.wis-tns.org}} Based on previous work, we use the observed
dispersion measure (DM) to bound the photon mass. In order to do so, we
estimate the redshifts of FRBs through package
FRUITBAT~\cite{Batten:2019bbf}\footnote{\url{https://fruitbat.readthedocs.io}}
and estimate the DM of host galaxies in a Bayesian framework. Also with the
help of a Bayesian method, we obtain a new photon mass bound, $m_{\gamma}
\leq 3.1\times 10^{-51}\rm\,kg \simeq 1.7 \times 10^{-15}\,eV/c^2$
($m_{\gamma} \leq 3.9\times 10^{-51}\rm\,kg \simeq 2.2 \times
10^{-15}\,eV/c^2$) at the $68\%$ $(95\%$) confidence level (C.L.), which
improves the previous bound of the photon mass from similar methods by a
factor of 2--3.

The paper is organized as follows. In the next section, we introduce a
theoretical framework for data analysis which includes a hypothesis for the
$\nu^{-2}$-behaved time delay, and a Bayesian framework to estimate the DM of
host galaxies and the photon mass. In Sec.~\ref{Sec:Result}, we present our
results on the estimation of the DM of host galaxies and the bound of
the photon mass from a combination of 129 FRBs. In
Sec.~\ref{Sec:Discussion} we present some discussions on the results.

\section{Theoretical framework}

In Sec.~\ref{Sec:hypothesis:of:time:delay} we introduce a hypothesis for
the $\nu^{-2}$-behaved time delay of FRBs in observation, and then utilize
a Bayesian framework in Sec.~\ref{Sec:Bayesian:analysis} to analyze the
observed FRB data. The analysis includes the estimation of the DM of host
galaxies and the bound of the photon mass.

\subsection{A hypothesis on the time delay}
\label{Sec:hypothesis:of:time:delay}

If photons have a non-zero rest mass, there will be a time delay in the
arrival times of FRB photons observed at different frequencies.
Observations of FRBs show an indisputable $\nu^{-2}$-dependent relation in
the time delay, $\Delta t \propto \nu^{-2}$~\cite{Petroff:2016tcr}. In our
calculation, we attribute the time delay to two causes: (i) the time delay
of electromagnetic waves caused by the propagation of photons through the
ionized medium, and (ii) the time delay caused by a non-zero rest mass of
photons.

On one hand, the astrophysical term, $\Delta t_{\rm DM}$, is caused by the
interaction between the ionizing medium and the electromagnetic waves. For
a photon with energy, $E=h\nu$, relative to a photon with an infinite
energy, the time delay $\Delta t_{\rm DM}$ is~\cite{2004hpa..book.....L},
\begin{equation}\label{Eq:Delta:t}
    \Delta t_{\mathrm{DM}}=\int \frac{\mathrm{d} l}{c_{0}}
    \frac{\nu_{\mathrm{p}}^{2}}{2 \nu^{2}} \simeq 4.15\,
    \mathrm{ms}\bigg(\frac{\mathrm{D}
    \mathrm{M}_{\mathrm{astro}}}{\mathrm{pc}\,
    \mathrm{cm}^{-3}}\bigg)\Big(\frac{\nu}{1\,
    \mathrm{GHz}}\Big)^{-2}\,,
\end{equation}
where the plasma frequency $\nu_{\mathrm{p}} \equiv \sqrt{n_{e} e^{2} / 4
\pi^{2} m_{e} \epsilon_{0}}$ with $e$ the charge of an electron, $n_e$ the
number density of electrons, $m_e$ the mass of an electron, and
$\epsilon_0$ the permittivity of free space. The definition of ${\rm
DM}_{\rm astro}$ is the integral of the electron number density along the
propagation path, $\mathrm{DM}_{\mathrm{astro}} \equiv \int n_{e}
\mathrm{d}l$~\cite{2004hpa..book.....L}. Considering a non-negligible
cosmological redshift, one has $\mathrm{DM}_{\mathrm{astro}} \equiv \int
{n_{e}^{(0)}}{(1+z)^{-1}} \mathrm{d} l$~\cite{Deng:2013aga}, where $z$ is
the redshift, and $n_{e}^{(0)}$ is the electron number density in the rest
frame.

There are multiple sources of $\rm DM_{astro}$, mainly including
contributions from the Milky Way (MW) galaxy, $\rm DM_{MW}$, from the
intergalactic medium (IGM), $\rm DM_{IGM}$, and from the host galaxy, $\rm
DM_{host}$. Therefore, the total DM contributed from the electromagnetic
wave propagating through the ionized medium can be written as,
\begin{equation}
	\rm DM_{astro}=DM_{MW}+DM_{IGM}+DM_{host} \,, 
\end{equation}
where we have included contributions from the near-source plasma (e.g.
supernova remnant, pulsars' wind nebula, HII region,
etc.~\cite{Yang:2017bls}) collectively in $\rm DM_{host}$. Generally,
people consider that the halo of the MW is mainly composed of dark matter
and there is no other evident interaction but gravity between dark matter
and electromagnetic waves. So we ignore the contribution from the MW halo to the total DM in our work~\cite{Navarro:1995iw,Moore:1999nt}.

On the other hand, comparing fiducial photons of infinite frequency with
Eq.~(\ref{Eq:velocity}), the time delay caused by the mass of the photon,
denoted as $\Delta t_{m_{\gamma}}$, is,
\begin{equation}\label{Eq:Delta:t_m_y}
    \Delta t_{m_{\gamma}}=\frac{1}{2 H_{0}}\bigg(\frac{m_{\gamma}
    c_{0}^{2}}{h \nu}\bigg)^{2} H_{\gamma}(z)\,,
\end{equation}
where the Hubble constant $H_{0}=67.66 \pm
0.42\,\mathrm{km}\,\mathrm{s}^{-1}\,\mathrm{Mpc}^{-1}$~\cite{Aghanim:2018eyx},
and $H_{\gamma}(z)$ is a dimensionless function of the source redshifit
$z$,
\begin{equation}\label{eq:H_gamma}
    H_{\gamma}(z) \equiv {\int_{0}^{z}}\frac{{\mathrm{d}
    z^{\prime}}}{{\left(1+z^{\prime}\right)^{2}}
    {\sqrt{\Omega_{\Lambda}+\left(1+z^{\prime}\right)^{3}
    \Omega_{\mathrm{m}}}}}\,,
\end{equation}
where the vacuum energy density $\Omega_{\Lambda}$ = $0.6889 \pm 0.0056$,
and the matter energy density $\Omega_{\mathrm m}$ = $0.3111 \pm 0.0056$~\cite{Aghanim:2018eyx}.

In our hypothesis, the total time delay can be written as, $\Delta t=\Delta
t_{\rm DM}+\Delta t_{m_{\gamma}}$, and these two time delays are both
$\nu^{-2}$-behaved which means that the total observed DM, $\rm DM_{obs}$,
can be written in a similar form, $\mathrm{DM}_{\mathrm{obs}} =
\mathrm{DM}_{\mathrm{astro}}+\mathrm{DM}_{\gamma}$, where the equivalent DM
corresponding to the time delay caused by the mass of photons
is~\cite{Shao:2017tuu},
\begin{equation}\label{Eq:DM_y}
    \mathrm{DM}_{\gamma} \equiv \frac{4 \pi^{2} m_{e} \epsilon_{0}
    c_{0}^{5}}{h^{2} e^{2}} \frac{H_{\gamma}(z)}{H_{0}} m_{\gamma}^{2}\,.
\end{equation}

\subsection{Bayesian framework}\label{Sec:Bayesian:analysis}

We work in a Bayesian framework developed in Ref.~\cite{Shao:2017tuu}. In
Bayesian analysis, the posterior distribution of parameter $\theta$ is
given as,
\begin{equation}\label{Eq:Bayesian:Equation}
    P\left(\theta \!\mid \mathcal{D}, \mathcal{H},
    \mathcal{I}\right)=\frac{P\left(\mathcal{D} \!\mid \theta, \mathcal{H},
    \mathcal{I}\right) P\left(\theta \!\mid \mathcal{H},
    \mathcal{I}\right)}{P(\mathcal{D} \!\mid \mathcal{H}, \mathcal{I})}\,,
\end{equation}
where $\mathcal{D}$ is data, $\mathcal{H}$ is the hypothesis given in
Sec.~\ref{Sec:hypothesis:of:time:delay}, $\mathcal{I}$ is all other
relevant background knowledge. $P\left(\mathcal{D} \!\mid \theta,
\mathcal{H}, \mathcal{I}\right)$ is the likelihood for the data,
$P\left(\theta \!\mid \mathcal{H}, \mathcal{I}\right)$ is the prior on the
parameter $\theta$, and $P(\mathcal{D}\!\mid \mathcal{H}, \mathcal{I})$ is
model evidence. With the prior and the likelihood, we can get the posterior
distribution of parameter $\theta$ from Eq.\,(\ref{Eq:Bayesian:Equation}).
In our work, we apply Bayesian analysis on the parameter estimation of $\rm
DM_{host}$ and $m_{\gamma}$ as follows. The model evidence merely plays a
normalization factor in our problem.

We have separated $\rm DM_{obs}$ into $\rm DM_{astro}$ and $\rm
DM_{\gamma}$, and will analyze each term in $\rm DM_{astro}$ to deduce the
likelihood of $\rm DM_{host}$.

Firstly, there are some electron distribution models with due uncertainties
for the MW from combining different observational results. The most widely
used models are the NE2001 model~\cite{Cordes:2002wz} and the YMW16
model~\cite{Yao_2017}. We mainly use the NE2001 model in our work. As for
the uncertainty estimation, we take the difference between the above two
models as a standard deviation. In principle, there are potential
contributions from the Galactic halo. As we discussed in
Sec.~\ref{Sec:hypothesis:of:time:delay}, considering that the contribution
from the Galactic halo is small and negligible in most cases, we do not
count it into $\rm DM_{obs}$. Even if there is a certain contribution, it
can be covered by our estimation of the error in $\rm DM_{MW}$. Excluding
$\rm DM_{MW}$, we can get the excess DM of the astrophysical term,
$\mathrm{DM}_{\mathrm{FRB}}^{\prime} \equiv \mathrm{DM}_{\mathrm{astro}} -
\mathrm{DM}_{\mathrm{MW}} = \mathrm{DM}_{\mathrm{host}} +
\mathrm{DM}_{\mathrm{IGM}}$.
	
Secondly, for the DM from IGM, $\rm DM_{IGM}$, we use a one-parameter model
to account for the scattering in the electrons from foreground
structures~\cite{Macquart:2020lln,MiraldaEscude:1998qs}. Its probability
distribution is,
\begin{align}\label{eq:P:IGM}
	P_{\rm IGM}( \left. \mathrm{DM}_{\rm IGM} \right| z )&=A
	\Delta^{-\beta} \exp \bigg[ -\frac{\left(\Delta^{-\alpha} -
	C_{0}\right)^{2}}{2 \alpha^{2} s^{2}}\bigg]\,, \\
	\Delta &\equiv \mathrm{DM}_{\rm IGM} /\left\langle\mathrm{DM}_{\rm
	IGM}\right\rangle > 0 \,,
\end{align}
where the normalization constant $A$ guarantees the integral of the
probability function to 1. The fractional fluctuation of the dispersion,
$s=Fz^{-0.5}$, is caused by the cosmological large-scale
structures which may exist in the foreground, and $F$ is used to quantify
the strength of the baryonic feedback. In Ref.~\cite{Macquart:2020lln}, the
authors selected $\alpha=3$ and $\beta=3$. The parameter $C_0$ can be
derived from the demand that the mean of the distribution of $\Delta$
satisfies $\langle\Delta\rangle=1$. Taking the average with respect to the
sky location, $\rm DM_{IGM}$ at redshift $z$ is,
\begin{equation}\label{Eq:DM_IGM_origin}
	\langle\mathrm{DM}_{\mathrm{IGM}}\rangle=\frac{3 c_{0} H_{0}
	\Omega_{\mathrm{b}} f_{\mathrm{IGM}}}{8 \pi G m_{\mathrm{p}}}
	\int_{0}^{z} \frac{g(z)\left(1+z^{\prime}\right) \mathrm{d}
	z^{\prime}}{\sqrt{\Omega_{\Lambda}+\left(1+z^{\prime}\right)^{3}
	\Omega_{\mathrm{m}}}}\,,
\end{equation}
where $m_{\rm p}$ is the proton mass, $\Omega_{\rm b}$ = $0.0487 \pm
0.0007$ is the baryonic matter energy density~\cite{Aghanim:2018eyx},
$f_{\rm IGM}\simeq 0.83$ is the baryonic fraction of IGM
\cite{Fukugita:1997bi}. In the electron mass fraction, $g(z) \equiv
\frac{3}{4} y_{1} \chi_{e, \mathrm{H}}(z)+\frac{1}{8} y_{2} \chi_{e,
\mathrm{He}}(z)$, the hydrogen and helium mass fractions are normalized to
their typical values $3/4$ and $1/4$ respectively. We have used $y_1\sim
1$, $y_2 \sim 1$, and $\chi_{e}$ is the ionized fraction in IGM, which for
hydrogen $\chi_{e,\rm H}(z) \simeq 1$ at $z<6$, and for helium, $ \chi_{ e,
\rm He}(z) \simeq 1$ at $z<2$. Therefore, we will only consider FRBs
with $z<2$, and we have~\cite{McQuinn:2008am},
\begin{equation}\label{Eq:DM_IGM}
	\langle\mathrm{DM}_{\mathrm{IGM}} \rangle=\frac{21 c_{0} H_{0}
	\Omega_{\mathrm{b}} f_{\mathrm{IGM}}}{64 \pi G
	m_{\mathrm{p}}}H_e(z)\,,
\end{equation}
\begin{equation}
	H_{e}(z) \equiv \int_{0}^{z} \frac{\left(1+z^{\prime}\right)
	\mathrm{d}
	z^{\prime}}{\sqrt{\Omega_{\Lambda}+\left(1+z^{\prime}\right)^{3}
	\Omega_{\mathrm{m}}}}\,,
\end{equation}
where we have rewritten the Hubble constant $H_0$ in a dimensionless
form, using $h_{70}=H_{0} /(70\,\mathrm{km}\, \mathrm{s}^{-1}\,
\mathrm{Mpc}^{-1})$.
	
For the last term in $\rm DM_{astro}$, we adopt a model for the probability
distribution of $\rm DM_{host}$. What is different from the above two terms
is that, unlikely the NE2001 model or Eq. (\ref{Eq:DM_IGM}), there is no
persuasive astrophysical model yet for $\rm DM_{host}$. So we take advantage of Bayesian analysis, and use a lognormal distribution model in our work~\cite{Macquart:2020lln}. There are two main considerations: (i) the lognormal distribution is a positive definite distribution, and (ii) it will have asymmetric tails at high values. These high values have many
potential origins like the gas local to the FRB (such as the HII region and the circumstellar medium) or the interstellar medium. These two
characteristics agree with the physical meaning of $\rm DM_{host}$.
Specifically, we use the probability distribution of $\rm DM_{host}$
as~\cite{Macquart:2020lln},
\begin{align}\label{eq:P:host}
	P_{\rm host}\big(\rm {DM}_{\rm host} \big| \mu, \sigma_{\rm
	host}\big)=&\frac{1}{(2 \pi)^{1 / 2} \mathrm{DM_{\rm host}}
	\sigma_{\rm host}}\nonumber\\
	&\times\exp \bigg[-\frac{\left(\log \mathrm{DM_{\rm
	host}}-\mu\right)^{2}}{2 \sigma_{\rm host}^{2}}\bigg],
\end{align}
where $\mu$, $\sigma_{\rm host}$ are the estimated parameters which are
used to calculate the expectation and variance of $\rm DM_{host}$
respectively. Note that in this equation we are assuming the standard
``${\rm pc\,cm}^{-3}$'' unit for DM.

Strictly speaking, the $\rm DM_{host}$ obtained in
Ref.~\cite{Macquart:2020lln} uses the assumption $m_{\gamma}=0$. It would
be ideal to allow for a non-zero photon mass. However, compared with $\rm
DM_{IGM}$, $\rm DM_{host}$ is not that large to make great difference on
the final result. So in this part, we temporally ignore the contribution of
the photon mass term, and use $\rm DM_{obs}$ as $\rm DM_{astro}$. This
treatment is tentatively feasible if we are putting a bound on the photon
mass, as we are doing it in this work, but will need to be refined if we
are to measure a non-zero $m_\gamma$.

\begin{table*}[h]
	\caption{Repeating FRBs that are used to estimate $\rm DM_{host}$.
	Sky position is given in Galactic longitude, $g_l$, and Galactic
	latitude, $g_b$. DM is in unit of $\rm pc~ cm^{-3}$, where $\rm
	DM_{obs}$ is from the fitting of the $\nu^{-2}$-behavior in the
	frequency-dependent time delay, and $\rm DM_{MW}$ is based on the
	NE2001 Galactic electron density model~\cite{Cordes:2002wz}. Redshift
	is given by the localization method.\label{Table:9:repeating:FRBs} }
	\centering
    \begin{tabular}{llllrrrr}
	    \toprule
	    	FRB&Telescope&$g_l$ (deg) &$g_b$ (deg) &$\rm DM_{MW}$&$\rm DM_{obs}$&$\sigma_{\rm DM}$&Redshift $z$\\
	    \midrule
	    	FRB 190711&ASKAP&310.9078&$-33.9023$&56.4&593.1&0.4&0.522\\
	    	FRB 190611&ASKAP&312.9352&$-33.2818$&57.83&321.4&0.2&0.378\\
	    	FRB 190608&ASKAP&53.2088&$-48.5296$&37.2&338.7&0.5&0.1178\\
	    	FRB 190523&DSA-10&117.03&44&37&760.8&0.6&0.66\\
	    	FRB 190102&ASKAP&312.6&$-33.49$&57.3&363.6&0.3&0.291\\
	    	FRB 181112&ASKAP&342.6&$-47.7$&102&589.27&0.03&0.4755\\
	    	FRB 180924&ASKAP&0.742467&$-49.414787$&40.5&361.42&0.06&0.3214\\
    	\bottomrule
    \end{tabular}
\end{table*}

We proceed to estimate the model likelihood by computing the likelihoods of
all FRBs and combining the $P_{\rm IGM}$ in Eq.~(\ref{eq:P:IGM}) and
$P_{\rm host}$ in Eq.~(\ref{eq:P:host}),
\begin{equation}
    \mathcal{L}_1=\prod_{i=1}^{N_{\mathrm{FRBs}}}
    P_{i}\Big(\mathrm{DM}_{\mathrm{FRB},i}^{\prime} \big| z_{i}\Big)\,,
\end{equation}
where $N_{\rm FRBs}$ is the number of FRBs with the index $i$ running over
FRBs. In the above equation, the likelihood for each FRB is
\begin{align}
    P_{i}(\mathrm{DM}_{\mathrm{FRB},i}^{\prime}&\!\mid
    z_{i})=\int_{0}^{\mathrm{DM}^{\prime}_{{\rm FRB},i}} P_{\rm
    host}\left(\mathrm{DM}_{\rm {host }} \!\mid\mu, \sigma_{\rm
    {host}}\right)\nonumber\\
    &\times P_{\rm {IGM}}\Big(\left.\mathrm{DM}_{\mathrm{FRB},
    i}^{\prime}-\mathrm{DM}_{\rm {host}} \right| z_{i}\Big)
    \mathrm{dDM}_{\rm {host}}\,.
\end{align}

Finally, we get a likelihood in a four-parameter space, ($\Omega_{\rm b}
h_{70},\ F,\ {\mu},\ \sigma_{\rm host}$), among which $\mu,\sigma_{\rm
host}$ are the (dimensionless) parameters that we need to calculate the
$\rm DM_{host}$. We choose the priors of each parameter to be uniform in
the following ranges: $\Omega_{\rm b}h_{70} \in [0.015,0.095]$, $F \in
[0.01,0.5]$, ${\rm e}^\mu \in [20,200]$, and $\sigma_{\rm host} \in
[0.2,2]$. \HW{}{The priors that we choose are based on reasonable consideration
and they were used in other literature as well.  If one has chosen other priors, as
long as that are not extreme, we do not expect significant changes to our
results.}

From discussions so far, we see that once we get $\rm DM_{host}$, with
above $\rm DM_{IGM}$ and $\rm DM_{MW}$, we can get $\rm DM_{astro}$ and
eventually deduce the value of $\rm DM_{\gamma}$ to bound $m_{\gamma}$. Now
we proceed to obtain the likelihood $\mathcal L_2$ for $m_{\gamma}$. As we
discussed, we can get $\rm DM_{MW}$ in the NE2001 model and the YMW16
model. We take the result of the NE2001 model as the central value and the
difference between the two models as the standard deviation. We take the
average $\rm DM_{IGM}$ as the expectation value using
Eq.\,(\ref{Eq:DM_IGM}), and the contribution of foreground structures is
covered in the uncertainty of $\rm DM_{IGM}$. We associate a 20$\%$
uncertainty to $\Omega_{\rm b}f_{\rm IGM}$ in the hope to cover the
inhomogeneity of IGM along different lines of sight. We use our result in
the parameter estimation of $\rm DM_{host}$ described earlier. To take the
cosmological evolution into account, we have multiplied it by a factor of
$(1+z)^{-1}$.

After estimating $\rm DM_{astro}$, we deduce $\rm DM_{\gamma}$ which is
related directly to $m_{\gamma}$. We apply a Bayesian framework to bound
the photon mass. Similarly, we adopt the posterior distribution,
\begin{equation}
    P\big(m_{\gamma}^{2} \big| \mathcal{D}, \mathcal{H},
    I\big)=\frac{P\big(\mathcal{D} \big| m_{\gamma}^{2}, \mathcal{H},
    I\big) P\big(m_{\gamma}^{2}\big|\mathcal{H}, I\big)}{P(\mathcal{D}
    \!\mid \mathcal{H}, I)}\,.
\end{equation}
Assuming independence of observations of different FRBs, we construct the
logarithm of likelihood as~\cite{Shao:2017tuu},
\begin{equation}\label{Eq:Likelihood:of:m_y}
    \ln \mathcal{L}_2=-\frac{1}{2} \sum_{i}
    \frac{\big(\mathrm{DM}_{\mathrm{obs}}^{i}-\mathrm{D}
    \mathrm{M}_{\mathrm{astro}}^{i}-\mathrm{D}
    \mathrm{M}_{\gamma}^{i}\big)^{2}}{\sigma_{i}^{2}}\,,
\end{equation}
where $\rm DM_{\gamma}$ is given in Eq.\,(\ref{Eq:DM_y}), $\rm DM_{astro}$
is composed of the above listed terms in our Markov-chain Monte Carlo
(MCMC) runs, $\sigma$ includes all uncertainties added in quadratic
(including uncertainties in $\rm DM_{obs}$, $\rm DM_{MW}$, $\rm DM_{IGM}$,
$\rm DM_{host}$, and the redshift $z$).

We adopt a uniform prior on $m_{\gamma}$ in the range $[10^{-69},10^{-42}]\rm\,
kg$. The uncertainty principle of quantum mechanics gives the lower bound since
the age of our Universe is finite, while the upper bound is chosen to satisfy
the demands in the approximation in Eq.\,(\ref{Eq:velocity}). This prior covers
a very wide range which we view as a conservative choice. \HW{}{In practice, for
a uniform prior, because the measure from 0 to $10^{-69}$\,kg is totally
negligible in any sense, if we have chosen the lower end at $m_\gamma = 0$ which
encloses the zero mass case in the Maxwell's theory, our results in the next
section stay practically the same.}

\begin{figure*}[htb]
	\centering
	\includegraphics[width=11cm]{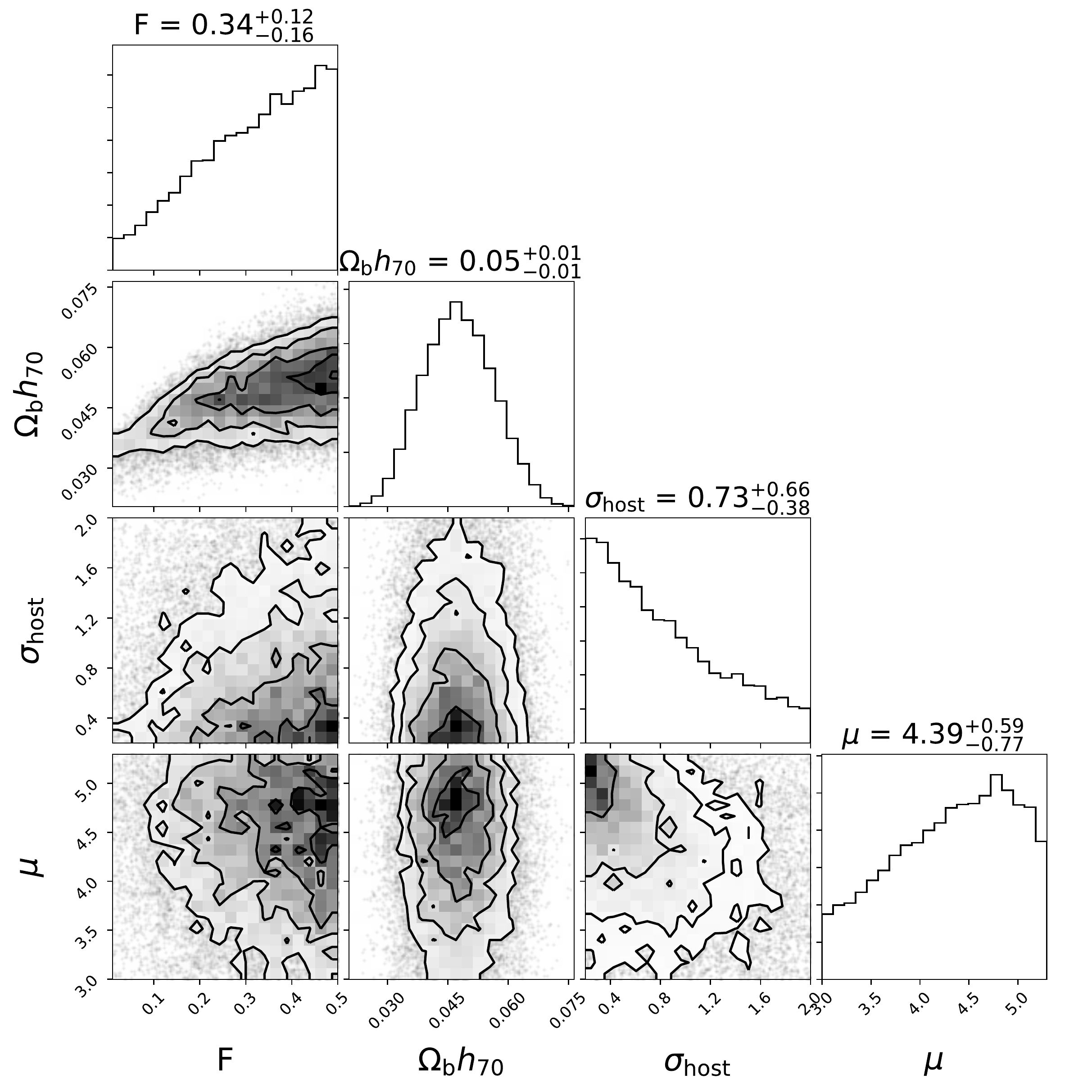}
	\caption{Parameter estimation posteriors for the DM of host galaxies in
	MCMC runs. The posteriors of $\mu$ and $\sigma$ are used to calculate the
	expectation and standard deviation of ${\rm DM}_{\rm host}$. 
	Notice that the four parameters in this figure are all dimensionless.
	\label{Fig:DM_host}}
	\end{figure*}

\begin{figure*}[htb]
	\centering
	\includegraphics[width=12cm]{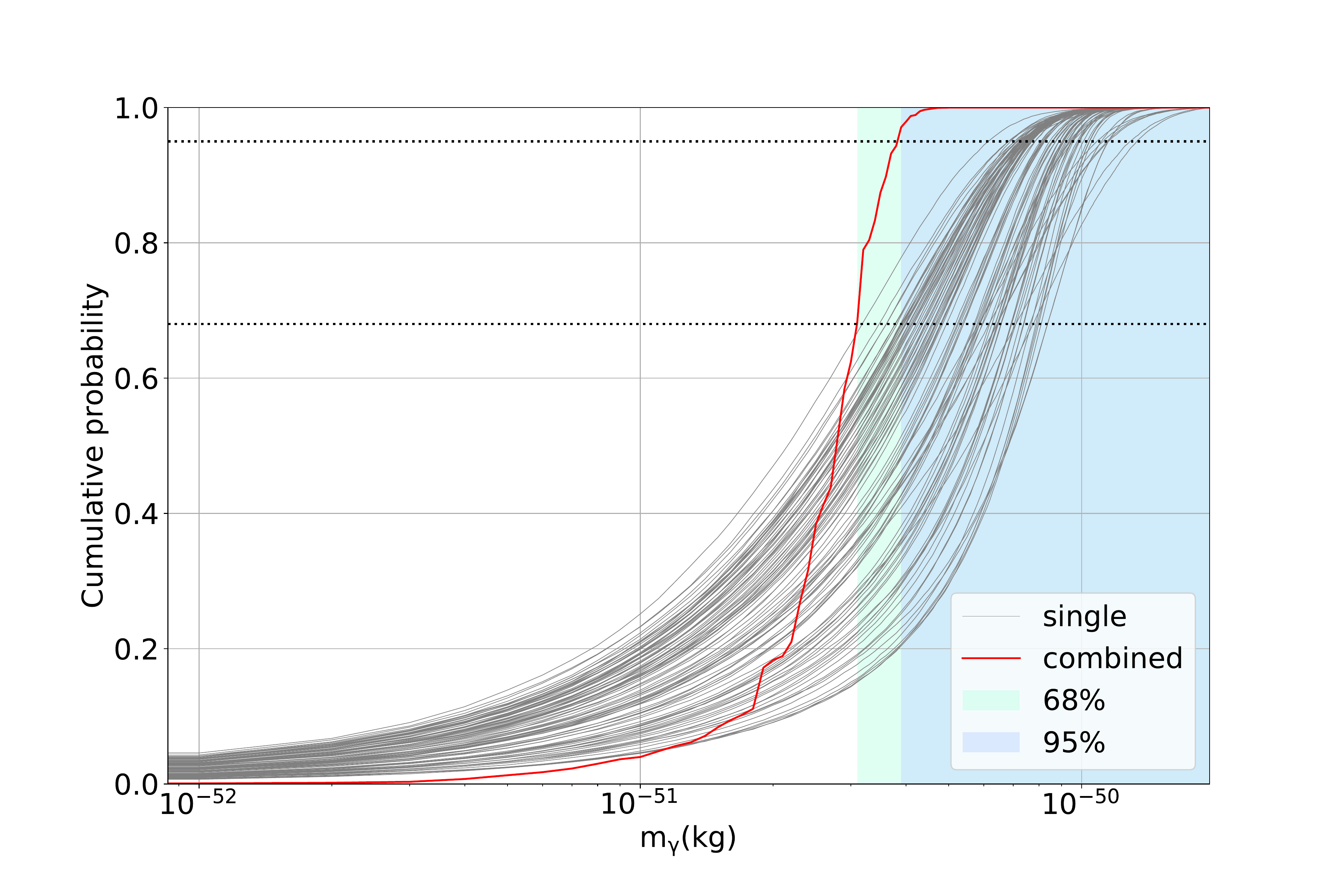}
	\caption{Cumulative posterior probability distributions for $m_\gamma$,
	with grey curves showing results from individual FRBs and the red curve
	showing the combined result. The excluded values for $m_{\gamma}$ from
	the combined result at 68$\%$ and 95$\%$ C.L.s are shown with shadowed
	areas.\label{Fig:m_y} }
\end{figure*}

\section{Results}\label{Sec:Result}

As mentioned, we use MCMC techniques to calculate the posterior of $\rm
DM_{host}$ and $m_{\gamma}$. We use PYTHON implementation of an
affine-invariant MCMC ensemble sampler~\cite{ForemanMackey:2012ig},
emcee,\footnote{\url{http://dan.iel.fm/emcee}} to perform simulations.

We use a PYTHON package, FRUITBAT~\cite{Batten:2019bbf}, to estimate the
redshifts of FRBs in the catalog. FRUITBAT is aimed to assist estimation of
redshifts, energies and the Galactic DM contributions of FRBs. Since
FRUITBAT only considers the line-of-sight $\rm DM_{MW}$ and $\rm DM_{IGM}$,
without the host galaxy and foreground structures that contribute to the DM
of many lines of sight, its result could be close to, but larger than, what
is given in the catalog already. Comparing the redshift determined by
direct measurements (e.g. from localization of the host galaxy for the
repeating FRBs) with that obtained by FRUITBAT~\cite{Petroff:2016tcr,
Chatterjee:2017dqg, Tendulkar:2017vuq,Marcote:2020ljw}, we choose $30\%$
uncertainty for these redshifts. For the repeating FRBs in the catalog, we
only use the first burst of each repeating FRB, because our calculation
require that every single burst is independent. Meanwhile, we use FRUITBAT
to estimate $\rm DM_{MW}$ in the NE2001 model and the YMW16 model.

There are several repeating FRBs observed in recent years. They represent
precious sources in studying the origin of FRBs. Before calculating $\rm
DM_{host}$, we first select sources with redshift measurement. We excluded
FRB~190614~\cite{Law2020ApJ...899..161L} and
FRB~20200120E~\cite{Bhardwaj2021arXiv210301295B} since their redshift
measurements are not that accurate and need more follow-up observations. We
also ignore FRB~121102~\cite{Chatterjee:2017dqg} and
FRB~180916~\cite{Marcote:2020ljw} owing to their low Galactic latitude. The
rest 7 repeating FRBs that we use are listed in
Table~\ref{Table:9:repeating:FRBs}. 

As stated in the last section, we use the NE2001 model and
Eq.\,(\ref{Eq:DM_IGM}) to estimate $\rm DM_{MW}$ and $\rm DM_{IGM}$
respectively. In our MCMC runs in estimating ${\rm DM}_{\rm host}$, there
are 30 chains with 15000 samples per chain after discarding the ``burn-in''
samples~\cite{ForemanMackey:2012ig}. In Fig.~\ref{Fig:DM_host}, we show the
samples returned by the MCMC sampler after discarding the ``burn-in''
samples. We choose our model for $\rm DM_{host}$ to follow a log-normal
distribution, characterized by an expectation $e^{\mu+\sigma^{2}/2}$ and a
median $e^{\mu}$. The standard deviation of the distribution is
$(e^{\sigma^{2}}-1)^{1/2}e^{\mu+\sigma^{2}/2}$. Since the standard
deviation given by MCMC is larger, we use the error of the expectation of
$\rm DM_{host}$ as the standard deviation in the subsequent calculation but
not the one calculated by the log-normal model. Our calculation gives
rather conservative results with $\rm DM_{host}=124\,pc\,cm^{-3}$, and
$\sigma=90\rm\,pc\,cm^{-3}$.

Up to March 15, 2021, there are 140 available sources in the FRB
catalog~\cite{Petroff:2016tcr}. We ignore FRBs whose errors of the observed
DMs are not given, and ones whose redshifts are larger than 2 in order to
satisfy the approximation in Eq.\,(\ref{Eq:DM_IGM_origin}). There are 129
FRBs left in our calculation. Ideally, we would calculate the
log-likelihood in the form of Eq.\,(\ref{Eq:Likelihood:of:m_y}) which means
simultaneously analyzing all FRB data in one run, but the computational
cost is quite high due to the large dimensionality of the parameter space.
Considering that FRBs are uncorrelated with each other, we combine the
individual bounds on $m_\gamma$ from the MCMC run of every single FRB. Such
an approach significantly reduces the computational burden. It is worth
noting that we discard weak bounds whose average values are larger than
$10^{-50.25}\,{\rm kg}$ because they give little contribution on the final
combined bound of $m_\gamma$ from all FRBs.

In our MCMC runs in estimating $m_\gamma$, there are 26 chains with 15000
samples per chain after discarding the burn-in samples. In
Fig.~\ref{Fig:m_y}, we show the accumulative posterior probability of
$m_{\gamma}$. From the figure, we obtain combined bounds on $m_{\gamma}$
from 129 FRBs,
\begin{eqnarray}
    m_{\gamma} &\leq 3.1\times 10^{-51}\rm\,kg \simeq 1.7 \times
    10^{-15}\,eV/c^2\,, \label{eq:68CL} \\
    m_{\gamma} &\leq 3.9\times 10^{-51}\rm\,kg \simeq 2.2 \times
    10^{-15}\,eV/c^2\,, \label{eq:95CL}
\end{eqnarray}
at 68\% C.L. and 95\% C.L. respectively.

\HW{}{It should be noted that we have ignored the contribution of $\rm{DM}_{\gamma}$
when we estimate the $\rm{DM_{host}}$ since the degeneracy of $\rm{DM_{host}}$
and $\rm{DM}_{\gamma}$ is very high. They are completely degenerate for a single
FRB.  However we want to stress that we obtained $\rm{DM_{host}}$ through a
Bayesian estimation of more than one hundred sources, and the degeneracy of
$\rm{DM_{host}}$ and $\rm{DM}_{\gamma}$ has been reduced to some degree, since
it is very unlikely for the Nature to {\it hide} a common value of the photon
mass for all FRBs through tuning the DM associated with each FRB. That is the
advantage of using a catalog of FRBs~\cite{Shao:2017tuu}. As we discussed in
Sec. 2.2, this method is statistically feasible because we are bounding the
photon mass instead of measuring a non-zero photon mass. In the future, if one
is at the stage to measure a non-zero photon mass, one needs to design more
sophisticated methods to overcome or reduce the degenecy between
$\rm{DM_{host}}$ and $\rm{DM}_{\gamma}$.}

\begin{figure*}[htb]
	\centering
	\includegraphics[width=10cm]{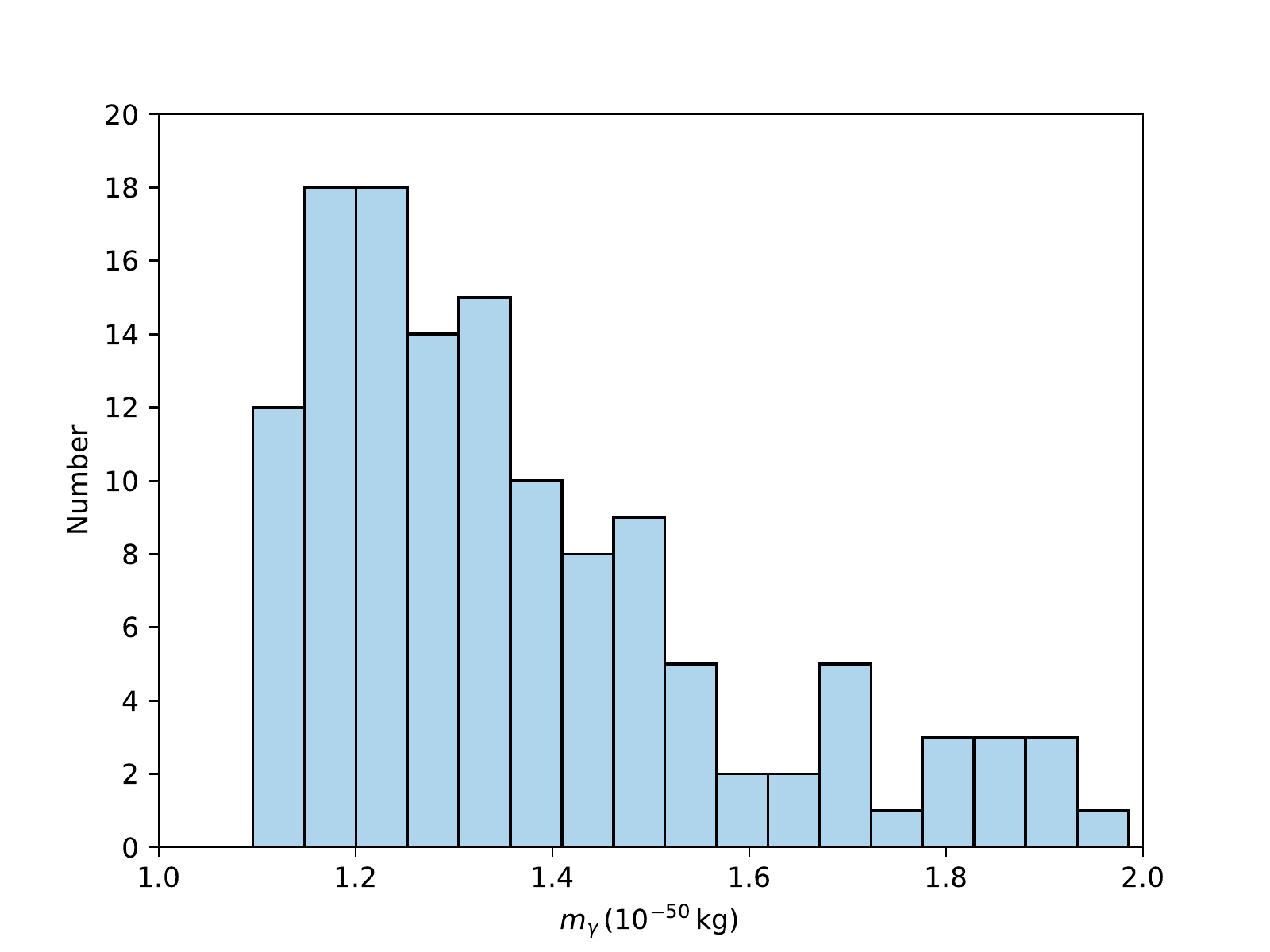}
	\caption{\HW{}{Histogram of the photon mass bounds provided by 129 FRBs base
	on the most conservative assumption that
	$\rm{DM_{astro}}=0$.}\label{Fig:hist_of_my} }
\end{figure*}

\HW{}{However, for a conservative consideration, we calculate the photon mass
bounds with selected FRBs by Eq.~(\ref{Eq:DM_y}) under the situation that
$\rm{DM_{astro}}=0$, in which the calculation is completely model-independent
and most conservative. The histogram of the obtained photon mass bounds is shown
in Fig.~\ref{Fig:hist_of_my}. Also, we read from the histogram  at 95\% C.L.\ a
very conservative bound for the photon mass, $m_\gamma \leq
1.9\times10^{-50}~\rm{kg}$.}

\section{Discussion}\label{Sec:Discussion}

Though Maxwell's theory is a well established theory in physics, it is
still intriguing to study its fundamentals to unprecedented precision due
to its tremendous importance in our everyday life and physical theories'
integrity. During past decades, there are various bounds on the photon mass
from {\it dynamic} tests. We give a brief overview for comparison with ours
in Eqs.~(\ref{eq:68CL}--\ref{eq:95CL}). In terrestrial laboratories,
physicists have tested the Coulomb law to great precision, and obtained a
photon mass, $m_\gamma < 1.6\times10^{-44}\,{\rm kg}$, in the Proca
theory~\cite{Williams:1971ms}. The Cavendish balance, which generates a
magnetic dipole vector potential moment with a suspended toroidal coil and
let the photon mass exhibit as the torque, gives a bound, $m_\gamma < \rm
1.2\times 10^{-48}\,kg$~\cite{Tu:2005ge,Lakes:1998mi,Luo:2003rz}. As
astrophysical observations reveal their potential in testing fundamental
physics~\cite{Tu:2005ge}, a few intriguing tests were conducted.
Magneto-hydro-dynamic phenomena of the Solar wind, using observations of
the flow pattern, give a bound $m_\gamma <\rm 1.5\times
10^{-48}\,kg$~\cite{Ryutov:2007zz}. Another interesting bound, $m_\gamma <
\rm 7.1\times10^{-56}\,kg$, comes from the so-called ``black hole bombs'',
where massive fields around rotating black holes trigger a superradiant
instability~\cite{Pani:2012vp}. We can see that the constraints given by
different tests vary greatly, spanning more than ten orders of magnitude.
As we stressed, these bounds are in general dependent on the underlying
massive photon theory.

In contrast, our limit in this paper is gained from a purely {\it
kinematic} test which is cleaner since it is generally quite independent of
the underlying massive photon theory. In the history of bounding the photon
mass by {\it kinematic} tests, pulsars were the earlier celestial objects
that were used~\cite{1969Natur.222..157W}. But even today, pulsars are mostly
observed in the MW which means that the distance of the propagation is
limited. With the discovery of $\gamma$-ray burst (GRB), it was found that
GRBs are at a cosmological distance, which makes them a better celestial
object than pulsars. In Refs.~\cite{Zhang:2016ruz, Schaefer:1998zg}, GRB
were used to bound $m_{\gamma}$, and the tightest bound is $m_\gamma < \rm
1.1\times10^{-47}\,kg$. Generically speaking, $\gamma$-ray photons are of
high energy, and GRBs are more suitable to test ultra-violet aspects of
theories, e.g. in testing the Lorentz symmetry of
spacetime~\cite{AmelinoCamelia:2008qg, Shao:2009bv, Shao:2010vj}. In
contrast, FRBs have low energy photons, and as emphasized earlier, they are
more suitable to test the photon mass.

In recent years, the ability in detecting FRBs has been greatly improved,
so that the potential in FRBs to constrain the photon mass has been
continuously explored (see Table~\ref{Table:Previous:bounds}). From the
study of the single burst such as FRB 121102 to the study of 9 repeating
bursts~\cite{Wei:2020wtf} or the study of subbursts of
FRB 121102~\cite{Xing:2019geq}, the research on FRBs is versatile in probing
the lower limit of the photon mass. In our work, we give the tightest bound
on the photon mass in {\it kinematic} tests so far with cosmological
propagation of FRBs in Eqs.~(\ref{eq:68CL}--\ref{eq:95CL}). 

Our
improvement is due to that, (a) more FRBs are used to bound the photon
mass, and (b) the redshift estimation done by the FRUITBAT
package~\cite{Batten:2019bbf}.

In the near future, the bound given by FRBs will be even tighter for
several reasons:
\begin{enumerate}[(i)]
    \item expected increment in the number of FRBs;
    \item more accurate measurement of the redshifts of FRBs;
    \item refinement in the knowledge about the origin of DMs, such as a
    better model for the electron distribution of the MW, a better
    understanding on the various astrophysical contributions to $\rm
    DM_{IGM}$ like the evolution of $f_{\rm IGM}$ which may correlate to
    the redshift~\cite{Walters:2019cie}; 
	\item more information on the host galaxy and FRBs themselves with
	improved analysis on the signal of FRBs and their Faraday rotation, and
	so on;
    \item and detection at lower frequency, which is advantageous in
    constraining $m_\gamma$ [see Eq.~(\ref{Eq:Delta:t_m_y})].
\end{enumerate}

With the increment in the number of FRBs, the population research comes to
reality. Through PYTHON packages such as
\emph{frbpoppy}~\cite{Gardenier:2019jit}, we hope to make predictions on
the detection capabilities of current and upcoming telescopes, and give
projected prediction on the bound of the photon mass. Such a study lays
beyond the scope of current paper.

\section*{Acknowledgements}

This work was supported by the National SKA Program of China
(2020SKA0120300), the National Natural Science Foundation of China
(11991053, 11975027, 11721303), the Young Elite Scientists Sponsorship
Program by the China Association for Science and Technology (2018QNRC001),
and the Max Planck Partner Group Program funded by the Max Planck Society.
It was partially supported by the Strategic Priority Research Program of
the Chinese Academy of Sciences (XDB23010200), and the High-performance
Computing Platform of Peking University. H.W.\ acknowledges the Principal's
Fund for the Undergraduate Student Research Study at Peking University.



\end{document}